\documentclass[reprint,aps,prapplied,twocolumn,amsmath,amssymb,nofootinbib,superscriptaddress]{revtex4-2}
\usepackage{graphicx,xcolor}
\usepackage{epstopdf}
\usepackage{xr}
\usepackage{hyperref}
\usepackage{cleveref,dsfont,bm}
\usepackage[export]{adjustbox}

\makeatletter
\newcommand*{\addFileDependency}[1]{
  \typeout{(#1)}
  \@addtofilelist{#1}
  \IfFileExists{#1}{}{\typeout{No file #1.}}
}
\makeatother

\newcommand{\bra}[1]{\langle#1|}
\newcommand{\ket}[1]{|#1\rangle}

\crefname{equation}{Eq.}{Eqs.}
\Crefname{equation}{Equation}{Equations}
\crefname{figure}{Fig.}{Figs.}
\Crefname{figure}{Figure}{Figures}
\crefname{section}{Sec.}{Secs.}
\crefname{subsection}{Subsec.}{Subsecs.}
\Crefname{section}{Section}{Sections}
\crefname{appendix}{Appendix}{Appendices}
\Crefname{appendix}{Appendix}{Appendices}

\DeclareMathOperator{\circleright}{\rotatebox[]{-90}{$\circlearrowright$}}

\DeclareMathOperator{\circleleft}{  \rotatebox[] {90}{$\circlearrowleft$}}

\newcommand{\IL}{{\rm IL}}
\newcommand{\IS}{{\rm IS}}
\newcommand{\R}{{\rm R}}

\begin{document}
\title[Article Title]
{Fano-enhanced low-loss on-chip superconducting microwave circulator}

\author{N. Pradeep Kumar}
\thanks{These authors contributed equally.}
\affiliation{Analog Quantum Circuits Pty.\ Ltd., Brisbane, QLD, Australia}
\affiliation{School of Mathematics and Physics, University of Queensland, Brisbane, QLD 4072, Australia}
\author{Dat Thanh Le}
\thanks{These authors contributed equally.} 
\affiliation{Analog Quantum Circuits Pty.\ Ltd., Brisbane, QLD, Australia}
\affiliation{School of Mathematics and Physics, University of Queensland, Brisbane, QLD 4072, Australia}

\author{Prasanna Pakkiam}
\affiliation{Analog Quantum Circuits Pty.\ Ltd., Brisbane, QLD, Australia}
\affiliation{School of Mathematics and Physics, University of Queensland, Brisbane, QLD 4072, Australia}
\author{Thomas M. Stace} 
\affiliation{Analog Quantum Circuits Pty.\ Ltd., Brisbane, QLD, Australia}
\affiliation{School of Mathematics and Physics, University of Queensland, Brisbane, QLD 4072, Australia}

\author{Arkady Fedorov}
\affiliation{Analog Quantum Circuits Pty.\ Ltd., Brisbane, QLD, Australia}
\affiliation{School of Mathematics and Physics, University of Queensland, Brisbane, QLD 4072, Australia}
\begin{abstract}
Ferrite-free circulators that are passive and readily integratable on a chip are highly sought-after in quantum technologies based on superconducting circuits. In our previous work, we implemented such a circulator using a three-Josephson-junction loop that exhibited unambiguous nonreciprocity and signal circulation, but required junction energies to be within $1\%$ of design values. 
This tolerance is tighter than standard junction fabrication methods provide, so we propose and demonstrate a design improvement that relaxes the required junction fabrication precision, allowing for higher device performance and fabrication yield.  Specifically, we introduce large direct capacitive couplings between the  waveguides to create strong Fano scattering interference.  We measure enhanced `circulation fidelity' above $97\%$, 
with optimised on-resonance insertion loss of $0.2$~dB, isolation of $18$~dB, and power reflectance of $-15$~dB, in good agreement with model calculations. 

\end{abstract}

\maketitle

\section{Introduction}

Circulators are a paradigmatic example of non-reciprocal devices with a wide use in telecommunication and microwave electronics \cite{Pozar11,Gu17,WangFerriteCirculatorPRAp21}. They are also indispensable for cryogenic microwave measurements where they are used to route weak microwave signals while protecting the system of interest from  thermal noise caused by higher temperature stages \cite{metelmannNonreciprocalPhotonTransmission2015,Ruesink16}.  However, conventional ferrite circulators are bulky and not compatible with microfabrication nor with superconducting circuits, and thus unsuitable for very-large-scale superconducting microwave networks. Given the drive to scale up superconducting quantum computers, designs for integrated microwave circulators on a chip are becoming critical \cite{reilly2015,Bravyi22TheFutureSuperconductingQuantumComputing}.

Many approaches to integrating circulators with other solid-state quantum circuits involve the application of strong magnetic fields, either real or synthesised with time-dependent control fields \cite{chapmanWidelyTunableOnChip2017, kamalNoiselessNonreciprocityParametric2011, kamalMinimalModelsNonreciprocal2017, estepMagneticfreeNonreciprocityIsolation2014, sliwaReconfigurableJosephsonCirculator2015, lecocqNonreciprocalMicrowaveSignal2017, fangGeneralizedNonreciprocityOptomechanical2017, metelmannNonreciprocalPhotonTransmission2015, petersonStrongNonreciprocityModulated2019, kerckhoffOnChipSuperconductingMicrowave2015, roushanChiralGroundstateCurrents2017, rosenthalBreakingLorentzReciprocity2017, AbdoJosephsonIsolatorNatCom19, AbdoDirectionalJosepshsonPRXQuantum21,PhysRevLett.121.123601,PhysRevLett.93.126804,reilly2017,mahoneyOnChipMicrowaveQuantum2017}. However, these approaches may also be incompatible with microfabricated superconducting systems, or add AC-control complexity.  

In contrast, our recent results demonstrated the  realisation of an on-chip superconducting circulator with only passive (i.e., DC) control~\cite{navarathnaPhysRevLett.130.037001, Fedorov2024}, based on a three-Josephson-junction loop \cite{kochTimereversalsymmetryBreakingCircuitQEDbased2010,mullerPassiveOnChipSuperconducting2018}. While nonreciprocity and microwave circulation were evident \cite{Fedorov2024}, the device performance was limited by asymmetry in Josephson junction energies.  In particular, the device  exhibited a `circulation fidelity' of $\sim80\%$ (i.e., the fidelity of the measured device scattering matrix relative to that of an ideal circulator) corresponding to an insertion loss of \mbox{2 dB}, when  post-selected on a specific quasiparticle configuration sector.

In this paper, we report the implementation of an improved design for a three-junction circulator device, which has been analysed in Refs.\ \cite{kochTimereversalsymmetryBreakingCircuitQEDbased2010,mullerPassiveOnChipSuperconducting2018,leOperatingPassiveOnchip2021,navarathnaPhysRevLett.130.037001,Fedorov2024}.  As in earlier work, the core of the device comprises three superconducting islands, indicated by different colours in \cref{fig:circuit}a that are connected to each other via Josephson junctions, and are  capacitively coupled to external waveguides. The key  advance reported here is the inclusion of  shunt capacitors that directly couple the waveguides  \cite{reilly2017,mahoneyOnChipMicrowaveQuantum2017,takeda2023passive}, indicated as $C_X$ in \cref{fig:circuit}a. This introduces an additional microwaves scattering pathway, giving rise to a Fano-like interference effect \cite{Miroshnichenko10FanoResonance,Fan03TemporalTheoryForFano}. 

Our theoretical simulations of the proposed design predict that high circulation fidelity, above $97\%$ (corresponding to an insertion loss of \mbox{$0.2$ dB}),  can be reached even when the spread in the Josephson energies is $\sim3\%$, which is achievable with standard electron beam lithography \cite{Kreikebaum20ImprovingwaferscaleJosephsonjunctionresistance,osmanSimplifiedJosephsonjunctionFabrication2021}. We confirm our theoretical predictions with experimental measurements, demonstrating significantly enhanced circulation performance.                                        

As with previously reported devices, our experimental system still suffers from significant quasiparticle hopping, so we use post-selection to characterise and optimise the device performance within  a single quasiparticle sector.  Relative to Ref.\ \cite{Fedorov2024}, we observe a ten-fold improvement in  device performance, with measured insertion loss of \mbox{$\rm{IL}=0.2$~dB}, isolation of \mbox{$\rm{IS}=18$~dB}, and power reflectance of \mbox{$\rm{R}=-15$~dB} at resonance.  

\section{Background}

\begin{figure}[t!]
    \centering
    \includegraphics[width=\columnwidth]{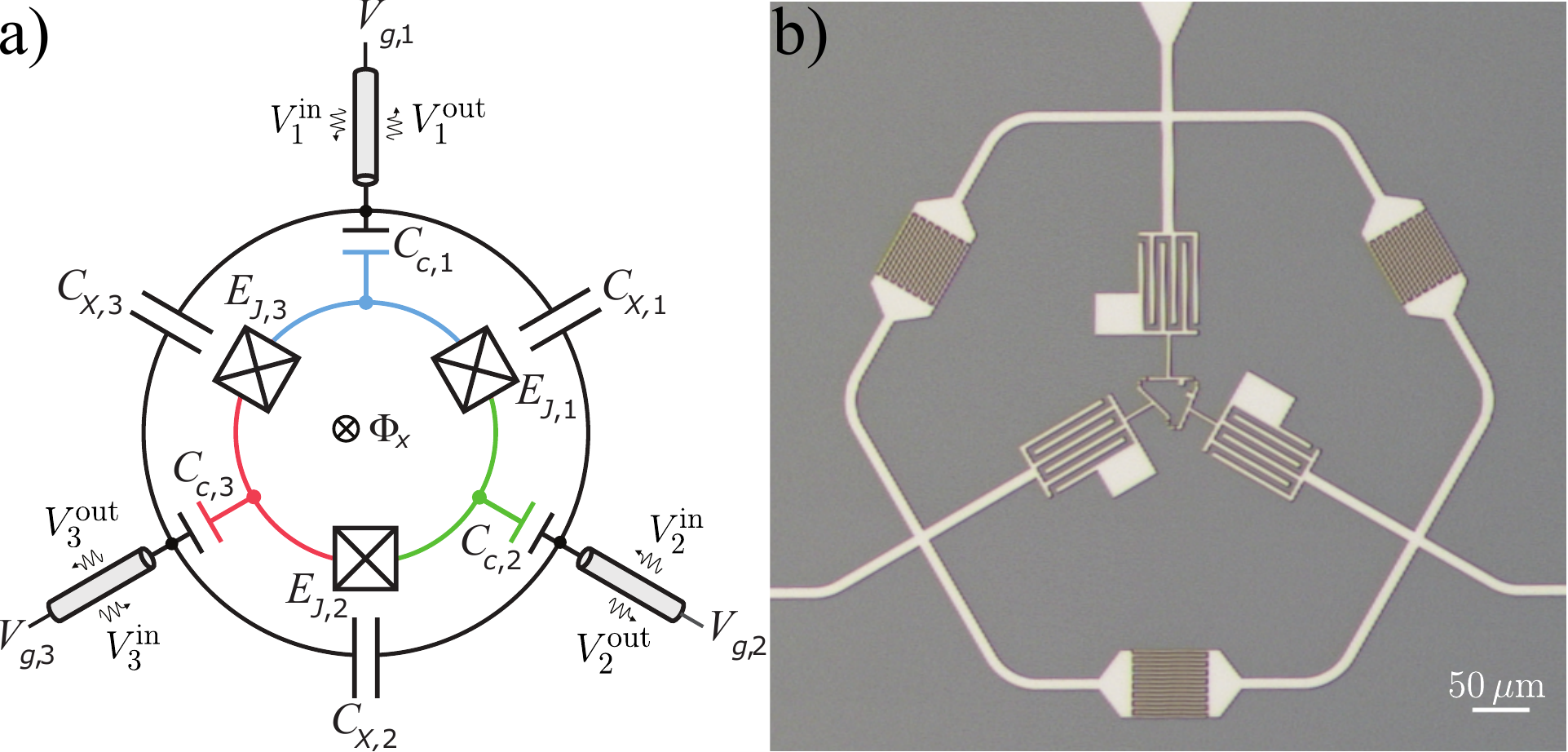}
    \caption{(a) Lumped-element circuit model of the capacitively-shunted circulator. The device consists of three Josephson junctions arranged in a loop that creates three superconducting islands (indicated by blue, green, and red colours). The islands are capacitively coupled to external waveguides via $C_C$, and the waveguides   are also directly  coupled to one-another by capacitive shunts $C_X$. 
    (b)  Optical microscope image of a fabricated device, where the centered triangle represents the loop formed by three Josephson junctions and the outer ring represents the inter-waveguide capacitances.}
    \label{fig:circuit}
\end{figure}

The electronic design of the capcitively-shunted circulator is shown in \cref{fig:circuit}a.    \Cref{fig:circuit}b implements this device, with the capacitive shunts  included in the hexagonal structure linking the  waveguides, which couple to off-chip signal sources and analysers.  The experimental setup is described in  detail in \cref{sec:exp}. 

To measure the scattering response of the system, we drive it with input voltage signals $V^{\mathrm{in}}_i$ and measure the output voltage signals $V^{\mathrm{out}}_j$ scattered by the device to determine the scattering matrix amplitudes $S_{ ji}=V^{\mathrm{out}}_j/V^{\mathrm{in}}_i$ with $i,j\in\{1,2,3\}$. We use an external DC flux bias threaded through the junction loop and DC charge biases applied to the superconducting islands ($\Phi_x$ and $V_{g,(1,2,3)}$ in \cref{fig:circuit}a) to control the operation of the device. 

Given the device's scattering matrix $S$, we quantify its circulation performance by defining the average clockwise and anticlockwise circulation fidelities, and the average reflection respectively as  
\begin{subequations}
\begin{eqnarray}
{\mathcal F}_{\,\circleright}&=&(|S_{13}|+|S_{32}|+|S_{21}|)/3,\\
{\mathcal F}_{\,\circleleft}&=&(|S_{12}|+|S_{23}|+|S_{31}|)/3,\\
{\mathcal R}&=&(|S_{11}|+|S_{22}|+|S_{33}|)/3.
\end{eqnarray}
\label{eqn:sums}%
\end{subequations}
An ideal clockwise circulator will have ${\mathcal F}_{\,\circleright}=1$ and \mbox{$\mathcal F_{\,\circleleft} = \mathcal{R}=0$} \cite{Fedorov2024}.

Our earliest experimental implementation of the three-junction-loop circulator (without shunt capacitors) was reported in \citet{navarathnaPhysRevLett.130.037001}, where we observed qualitative nonreciprocal behaviour, i.e.\ \mbox{$S_{ij}\neq S_{ji}$} for $i\ne j$, albeit with a low circulation fidelity $\sim 50 \% $. Based on detailed modelling, we attributed this low circulation performance to electrical asymmetry in the Josephson junctions, which had a large relative spread in Josephson energies $\delta E_J \sim 8.5\%$, where  \mbox{$\delta E_J \equiv (\mathrm{max}_j[E_{J,j}] - \mathrm{min}_j [E_{J,j}] )/\bar E_{J,j}$}. 

This stems from the fact that a large value of $\delta E_J$ determines the frequency splitting between the nearly degenerate first and second excited states of the loop. The interference of scattering pathways mediated by these states is responsible for signal circulation in the device \cite{leOperatingPassiveOnchip2021}, so that when this splitting is much larger than the waveguide coupling strength, an external drive cannot simultaneously couple strongly to both of the excited states, thus limiting the nonreciprocal interference.

Furthermore, the device reported in  Ref.\ \cite{navarathnaPhysRevLett.130.037001} was sensitive to charge fluctuations. In particular, quasiparticle tunneling between the superconducting islands created four quasiparticle sectors, each with a distinct scattering response \cite{leOperatingPassiveOnchip2021}. The average quasiparticle lifetime of the device in Ref.\ \cite{navarathnaPhysRevLett.130.037001} was found to be $\tau^{\rm{(qp)}}\sim 200\,  \mu\mathrm{s}$.  

The second iteration of the three-junction-loop circulator reported in \citet{Fedorov2024} (also without shunt capacitors)  featured a more geometrically symmetric design with a smaller spread in Josephson energies $\delta E_J\sim 2.2\%$. Additional design optimisation ensured that the capacitance matrix of the system was electrically symmetric. With these improvements, the device showed significant nonreciprocity and reached circulation fidelity $\sim 80\%$ (post-selected over quasiparticle sectors), for both clockwise and counter-clockwise circulation. The average quasiparticle lifetime was also improved to $\tau^{\rm{(qp)}}\sim 4\, \mathrm{ms}$, due to enhanced infrared shielding and a change in electronic parameters to reduce the charge-parity-switching rates.

To attain higher circulation performance in the three-junction system investigated in Ref.\ \cite{Fedorov2024}, modelling predicted that we need $\delta E_J\lesssim 1\%$ 
\cite{leOperatingPassiveOnchip2021}, which is more constrained than  standard junction fabrication precision allows for.  Instead, this level of precision can only be reliably assured with  post-fabrication treatments such as laser annealing \cite{hertzbergLaserannealingJosephsonJunctions2021}. 

In the following sections, we demonstrate an alternative approach to improving the circulation performance, based on Fano interference arising from the inclusion of  shunt capacitors.  We first develop a detailed theoretical model of the system, and use this to show that including relatively large waveguide shunt capacitors in the circulator design  relaxes the  required  junction fabrication precision, so that circulation becomes more robust against variations in Josephson junction energies. 
We then implement the device design, shown in \cref{fig:circuit}b, to demonstrate the experimental performance of the system, confirming good agreement between theory and experiment, as well as high quality circulation, after accounting for quasiparticle noise.

\section{Modelling and Analysis}

In this section, we develop a theoretical input-output model, which we use to show that introducing direct capacitive couplings between the waveguides, as shown in \cref{fig:circuit}a, enhances the circulation fidelity close to the ideal, even when asymmetry in Josephson energies is relatively large. 

\subsection{SLH master equation}

We derive a master equation for a three-junction loop capacitively coupled to input-output waveguides, including direct waveguide shunt capacitors mutually coupling the waveguides. We begin by first describing the bare Hamiltonian of the junction loop \cite{kochTimereversalsymmetryBreakingCircuitQEDbased2010, mullerPassiveOnChipSuperconducting2018, leOperatingPassiveOnchip2021}
\begin{eqnarray}
   \hat H_{\rm loop} &=& E_{C_{\Sigma}} \big( (\hat n'_1 - \tfrac{1}{2}(n_0 +n_{g,1}- n_{g,3}))^2 \nonumber  \\
&& + (\hat n'_2 + \tfrac{1}{2}(n_0 +n_{g,2}- n_{g,3}))^2 -\hat n'_1 \hat n'_2 \big) \nonumber \\
&& -E_{J,1} \cos(\hat \phi'_1 - \tfrac{1}{3}\phi_x) - E_{J,2}\cos(\hat \phi'_2- \tfrac{1}{3}\phi_x) \nonumber \\
&& -E_{J,3} \cos(\hat \phi'_1 + \hat \phi'_2  + \tfrac{1}{3}\phi_x)  \label{eq:LoopHamiltonian},
\end{eqnarray}
which depends on a single charging energy $E_{C_{\Sigma}}$ (under the assumption that the system capacitances are symmetric), three Josephson energies $E_{J,j}$, three charge biases $n_{g,j}$, and a flux bias $\phi_x$. In Eq.\ \eqref{eq:LoopHamiltonian}, the charge operators $\hat n'_1$ and $\hat n'_2$, the conversed total charge $n_0$, and the phase (difference) operators $\hat \phi'_1$ and $\hat \phi'_2$ are related to the original charge and phase operators of the superconducting islands as follows, $\hat n'_1 = \hat n_1$, $\hat n'_2 = -\hat n_2$, $n_0 = \hat n_1 + \hat n_2 + \hat n_3$, $\hat \phi'_1 = \hat \phi_1 - \hat \phi_3$, and $\hat \phi'_2 = \hat \phi_3 - \hat \phi_2$. 
The form of the kinetic (charging) energy in Eq.\ \eqref{eq:LoopHamiltonian} implies that the loop Hamiltonian $\hat H_{\rm loop}$ actually depends on the relative bias charges between the islands, $n_{g,1} - n_{g,3}$ and $n_{g,2} - n_{g,3}$. Therefore, tuning two charge biases only, e.g., $n_{g,1}$ and $n_{g,2}$, as well as the flux bias $\phi_x$ suffices to control the total operation of the circulator device. In terms of its eigenbasis \mbox{$\{ \ket{k}; k=0,1,2, \dots \}$}, $\hat H_{\rm loop}$ is expressed as $\hat H_{\rm loop} =   \sum_{k\geq 1} \hbar \omega_k \ket{k}\bra{k}$,
where $\omega_k$ is the loop transition frequency from the ground state $\ket{0}$ to the excited state $\ket{k}$ ($k\geq 1$).   

We consider injecting single-mode weak coherent fields with coherent amplitudes $\langle\hat{\boldsymbol{a}}^{\rm in}\rangle = (\alpha_1, \alpha_2, \alpha_3)^{\intercal}$ at a drive frequency $\omega_{\rm d}$ to the three waveguide ports. Following the `Scattering-Lindblad-Hamiltonian' (SLH) formalism \cite{Josh17SLHFramework}, we model the coherent input fields, the coupled waveguides, and the junction loop as three cascaded systems with their SLH triples 
respectively given by
\begin{eqnarray}
    G_{\rm d} &=& (\mathds 1_{3\times 3}, \hat{\bm L}_{\rm d},0), \\
    G_{\rm wg} &=& ({\bf A}, 0,0 ), \label{eq:WaveguideSLHtriple} \\
    G_{\rm loop} &=& (\mathds 1_{3\times 3}, \hat{\bm{L}}_{\rm loop}, \hat H_{\rm loop}),
\end{eqnarray}
\noindent where $ \hat{\bm L}_{\rm d} = (\alpha_1 \hat{  \mathds I}, \alpha_2 \hat{\mathds I}, \alpha_3 \hat {\mathds I})^\intercal$ denotes the coupling operators associated with the input drive fields and \mbox{$\hat {\bm{L}}_{\rm loop} = (\sqrt{\Gamma} \hat q_{1,-}, \sqrt{\Gamma} \hat q_{2,-}, \sqrt{\Gamma} \hat q_{3,-} )^\intercal$} denotes the coupling operators associated with the junction loop. Here $\Gamma$ is the coupling strength between the junction loop and the waveguides and $\hat q_{j,-} = \sum_{k<\ell} \langle k | \hat q_j | \ell \rangle \ket{k}\bra{\ell} $ are the upper triangularised parts (in the junction loop eigenbasis) of the operators $\hat q_j$, where $\hat q_1 = \hat n'_1, \hat q_2 = -\hat n'_2,$ and $\hat q_3 = - \hat n'_1 + \hat n'_2$.

$G_{\rm d}$ and $G_{\rm loop}$ represent three-port systems, while $G_{\rm wg}$ represents a six-port scattering system, with three `exterior' and three `interior' ports; its $6\times 6$ scattering matrix $\bf A$ is derived in \cref{append:6x6WaveguideSmatDerivation}. As analysed in \cref{append:SLHWithFeedback}, 
the `interior' ports of $G_{\rm wg}$ are connected to the ports of $G_{\rm loop}$ in a feedback configuration. We therefore apply both the SLH series and feedback rules \cite{Josh17SLHFramework} to cascade the total drive-waveguide-loop SLH triple
\begin{equation}
    G_{\rm tot} = G_{\rm d} \triangleleft (G_{\rm wg} 
    \hookleftarrow
    G_{\rm loop}),
\label{eq:TotalSLHTriple}
\end{equation}
which collectively describes a three-port device.  Here we have introduced a new SLH composition notation, \mbox{$A\hookleftarrow B$} to indicate that systems $A$ (outer) and $B$ (inner) are coupled in a feedback loop. \Cref{eq:TotalSLHTriple} is the basis for the SLH modelling, and further details are provided in \cref{append:SLHWithFeedback}, and  illustrated  in \cref{fig:scattering}c.  

Given $G_{\rm tot}$, the Linblad master equation for the density operator $\rho$ of the loop system in a frame rotating at the drive frequency $\omega_{\rm d}$ is given by
\begin{equation}
    \dot \rho = - i [\hat H'_{\rm tot}, \rho ] + \sum_{j=1}^3 \mathcal D [\hat a^{\rm out}_j] \rho, \label{eq:SLHme}
\end{equation}
where $\mathcal D[\hat o] \rho = \tfrac{1}{2} (2 \hat{o} \rho \hat{o}^\dag - \rho \hat{o}^\dag \hat{o} - \hat{o}^\dag {\hat o} \rho ) $ and
\begin{align}
    \hat H'_{\rm tot} &= \hat H'_{\rm loop} + \hat H_{\rm s} + \hat H_{\rm d}, \label{eq:HtotRotatingFrame} \\
    \hat{\bm a}^{\rm out} &= \hat{\bm L}_{\rm w \hookleftarrow l} + S_{\rm w \hookleftarrow l} \hat{\bm L}_{\rm d}. \label{eq:InputOutput}
\end{align}
Here $\hat H'_{\rm loop} = \sum_{k \geq 1} (\omega_k - \omega_{\rm d}) \ket{k} \bra{k}$, $\hat H_{\rm s}$ and $\hat H_{\rm d}$ represent the frequency shifts and the driving fields to the junction loop system, $\hat {\bm a}^{\rm out} = (\hat a^{\rm out}_1, \hat a^{\rm out}_2, \hat a^{\rm out}_3)^\intercal$ denotes the output fields, and $\hat {\bm L}_{\rm w\hookleftarrow l}$ and $S_{\rm w\hookleftarrow l}$ are the coupling operators and the scattering matrix of the feedback-reduced cascaded system $G_{\rm wg} \hookleftarrow G_{\rm loop}$. Explicit expressions of the operators in Eqs.\ \eqref{eq:HtotRotatingFrame} and \eqref{eq:InputOutput} are given in \cref{append:SLHWithFeedback}. Since 
\mbox{$V^{\mathrm{in}}_j = K \alpha_j$} and \mbox{$V^{\mathrm{out}}_j = K \langle \hat{a}^{\mathrm{out}}_j \rangle$}, where $K$ is a conversion factor and $\langle \hat {\mathcal{O}} \rangle = \mathrm{Tr} (\hat {\mathcal O} \rho) $, the scattering matrix $S$ with its elements \mbox{$S_{ji} = V^{\mathrm{out}}_j/V^{\mathrm{in}}_i = \langle \hat a^{\mathrm{out}}_j \rangle /\alpha_i  $} can be computed  numerically from \cref{eq:SLHme} and  \cref{eq:InputOutput}, to obtain $\langle \hat a^{\mathrm{out}}_j \rangle$.

\begin{figure}[t!]
    \centering
    \includegraphics{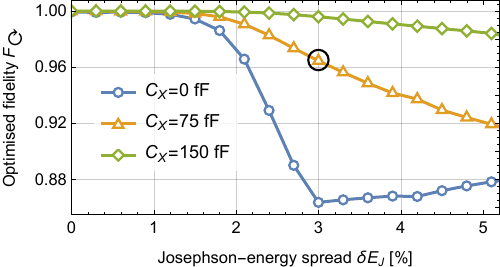}
    \caption{ Optimised circulation fidelity, ${\mathcal F}_{\,\circleright}$, as a function of the Josephson-energy spread  $\delta E_J$ for increasing values of the inter-waveguide capacitances, $C_X =\{0, 75, 150\}\, \rm fF$. The fidelity is computed from the scattering matrix $S$ computed with the master equation in \cref{eq:SLHme}. The junction loop parameters are $E_{C_{\Sigma}}/h = 3.09 \, \rm{GHz}$ and $E_{J,2} = 15.03  \, \mathrm{GHz}$, and we allow variations in  $E_{J,1/3}$ as $E_{J,1} = E_{J,2} (1-\delta E_{J}/2) $, and $E_{J,3} = E_{J,2} (1+\delta E_{J}/2)$. The black circle marks the point where $\delta  E_J = 3\%$, $C_X=75\, \rm fF$, yielding an optimised fidelity ${\mathcal F}_{\,\circleright} \gtrsim 97\%$, which is consistent with the measured circulation fidelity in \cref{fig:S-matrix}. 
    }
    \label{fig:fideVsEJasym}
\end{figure}

\subsection{Reduced sensitivity to junction asymmetry}

\begin{figure}[t!]
    \centering
\includegraphics[width=0.95\columnwidth]{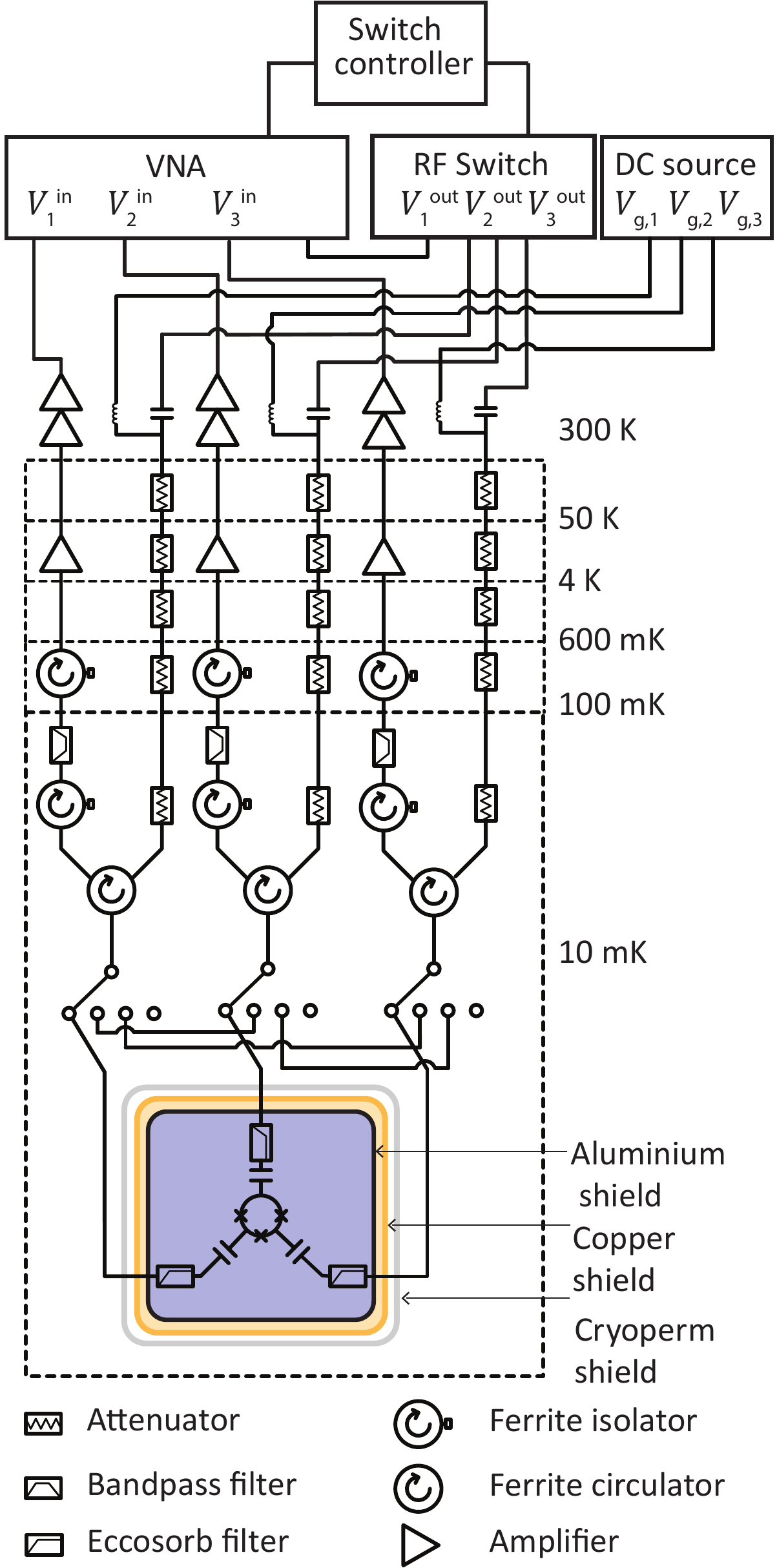}
    \caption{Circuit diagram of the experimental setup inside the dilution refrigerator. The ports of the device are connected to radial microwave switches to allow for bypassing the device and calibrating the input and output lines. The inputs $V^{\rm in}_j$ are sent through attenuators to the
    device. DC voltages $V_{g,j}$ are added to the RF input lines via bias
    tees at room temperature. Each of the outputs $V^{\rm out}_j$ from the device
    goes through an eccosorb filter, a circulator, two isolators, a band pass filter, a HEMT amplifier, and room temperature amplifiers. }
    \label{fig:Diagram}
\end{figure}

The inclusion of the waveguide shunt capacitors modifies the scattering mechanism in the proposed circulator. In particular, in \cref{append:AdiabaticElimination} we  adiabatically eliminate the loop degrees of freedom to approximate the system scattering matrix, at low drive powers, as \mbox{$S =  ( {\mathds 1} + R_{\rm loop}).S_{\rm w \hookleftarrow l}$, where $R_{\rm loop}$} represents the response of the junction loop to the external drives. 
This shows that scattering of the whole capacitively-shunted-waveguide  and junction-loop system is mediated via two pathways, namely, a direct pathway represented by the waveguide scattering $S_{\rm w \hookleftarrow l}$ and the loop-mediated scattering  represented by $R_{\rm loop}$. This gives rise to the Fano interference effect with asymmetric line shapes in the transmission and reflection spectra \cite{Miroshnichenko10FanoResonance,Fan03TemporalTheoryForFano}.

In what follows, we provide numerical evidence that the waveguide shunt capacitors reduce the sensitivity of the circulation fidelity to variations in the Josephson junctions fabricated in the loop. To this end, we numerically solve the master equation in \cref{eq:SLHme}, compute the output fields, and determine the scattering matrix $S$.

For simulation purposes, we take 
$E_{C_{\Sigma}}/h = 3.09 \, \rm GHz$, \mbox{$E_{J,2}/h \equiv E_{J}/h  =15.03\, \mathrm{GHz}$} to match  experimental parameters discussed in \cref{sec:exp}, following, and we allow $E_{J,1/3}$ to vary as
\mbox{$E_{J,1} = E_{J} (1-\delta E_{J}/2) $}, and \mbox{$E_{J,3} = E_J (1+\delta E_{J}/2) $}, where $\delta E_{J}$ is the fractional spread in junction energies. 
For  \mbox{$0\leq \delta E_{J}\leq5\%$}, we numerically optimise the flux and charge biases, and the drive frequency to find the optimal clockwise circulation fidelity at each $\delta E_{J}$.  This `maximally asymmetric' choice of junction energies for $E_{J,1/3}$ relative to the mean value $E_J$ represents the worst-case-scenario circulation fidelity \cite{leOperatingPassiveOnchip2021}, so provides an estimate of the worst-case effect of asymmetry on the circulation performance.

The simulation results are shown in \cref{fig:fideVsEJasym}, where we plot the optimised circulation fidelity, ${\mathcal F}_{\,\circleright}$, as a function of the Josephson-energy spread $\delta E_J$ for three values of the inter-waveguide capacitances \mbox{
$C_X \in \{ 0, 75, 150 \}\, \mathrm{fF}$}. We observe that for small junction spread, the fidelity is high, i.e., ${\mathcal F}_{\,\circleright}\approx1$ for \mbox{$\delta E_J \lesssim 1\%$}, regardless of the value of $C_X$.  For larger asymmetry, the fidelity 
improves with increasing  $C_X$.  In particular, the curve for $C_X = 75\, \mathrm{fF}$, which corresponds to our experimental value,  shows high circulation performance up to $\delta E_J \sim 3\%$.
These simulations indicate that large inter-waveguide capacitances substantially enhance the robustness of the circulation fidelity against Josephson-junction asymmetry.

\section{Experimental results}
\label{sec:exp}

\subsection{Device fabrication and measurement}

The device shown in \cref{fig:circuit}b was composed of four layers of aluminium deposited on a high resistivity silicon wafer with different thicknesses to reduce quasiparticle tunneling by gap engineering~\cite{kalashnikovBifluxonFluxonParityProtectedSuperconducting2020}. The first layer of 100 nm formed the capacitors and the basic structure of the junction loop. Standard double-angle evaporation was then used to deposit two layers of aluminium, of \mbox{20 nm} and 60 nm respectively, with a single oxidation step between the two aluminium deposition stages to form three Josephson junctions. 
After evaporation, the chip was diced and bonded on a holder suitable for cryogenic measurements in a dilution refrigerator operating at a base temperature of \mbox{10 mK}.

The  device was characterised using a fast, room-temperature microwave switch to sequentially direct the external drive to each of the three input ports for 100 $\mu$s, and measuring the response at the three output ports with a vector network analyser (VNA), as shown in \cref{fig:Diagram}.

\subsection{Spectral response}

We first measured the spectral response of the system as a function of the magnetic flux in the junction loop provided by a small external coil mounted on the bottom of the sample holder. 
In \cref{fig:Spectrum}, the measured spectrum, which features a characteristic $\textsf Y$-shape \cite{mullerPassiveOnChipSuperconducting2018,leOperatingPassiveOnchip2021,navarathnaPhysRevLett.130.037001,Fedorov2024}, shows good agreement with the theoretical model,  and the fitting provides an estimate for the electronic parameters of the device. 

Specifically, the model fitting returns an on-site charging energy \mbox{$E_{C_{\Sigma}}/h=(2e)^2/(h C_{\Sigma})=3.09$ GHz}, which corresponds to a total island capacitance of \mbox{$C_\Sigma=C_{g}+C_C+3C_J=37.5$ fF}, where $C_g$, $C_C$, and $C_J$ are the ground, waveguide-loop coupling, and junction capacitances respectively. This is close to the design value of \mbox{$C_\Sigma=40$ fF}, from  designed capacitance values \mbox{$C_g = 2.3$ fF},  \mbox{$C_C=27.5$ fF}, and \mbox{$C_J=3.5$ fF}. The fitted values for the Josephson energies are \mbox{$E_{J,(1,2,3)}/h=\{14.73, 15.15,15.22\}\, \rm GHz$} with a spread $\delta E_J=3.2\%$. These values are reasonably consistent with the measured room-temperature junction resistances
\mbox{$R_{J,(1,2,3)}=\{6.86,7, 7.02\}\, \mathrm{k}\Omega$}  \cite{Ambegaokar1963}, noting that junction resistances may drift slightly, relative to one-another, over several days.

\subsection{Scattering analysis}

We next measured the full $3\times3$ complex-valued scattering matrix $S$.

\begin{figure}[t!]
    \centering
\includegraphics{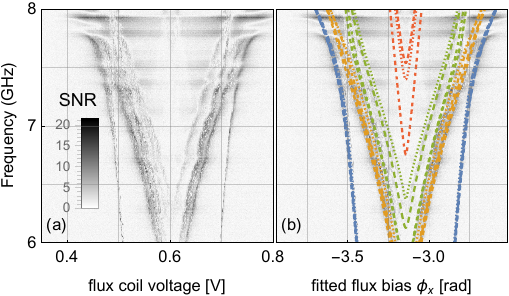}
    \caption{(a) Spectral response for voltage scattering from port 1 to port 2 while sweeping over the flux coil voltage. The raw voltage output at each frequency is scaled so that the far-off-resonant background data (i.e.\ where the flux coil voltage is $<0.4\textrm{ V}$) has zero mean and unit variance; the gray-scale thus represents the output signal-to-noise ratio of $|S_{21}|$. (b)  Fitted model spectral lines superimposed on the measured spectrum, including four distinct quasiparticle sectors; the model spectrum obscure the experimental features across the flux and frequency scan, indicating high quality agreement between theory and experiment. The horizontal features in the data are attributed to weak resonances from reflections in the cabling. The fitting process also converts the dimensional flux coil voltage into a dimensionless flux bias $\phi_x = 2\pi \Phi_x/\Phi_0$.
    Different colours correspond to transition frequencies from the ground state to the first excited state (blue), second (orange), third (green), and fourth (red). The fourfold multiplet within the fitted spectrum is due to  four  quasiparticle sectors.
 }
    \label{fig:Spectrum}
\end{figure}

The raw data obtained at the VNA includes the response of the scattering matrix $S$ of the capacitively-shunted on-chip circulator, along with the attenuation, amplification, and loss inside the cabling in the dilution refrigerator.  To isolate the device scattering matrix, $S$, we use a two-step calibration procedure. 
In the first step, we introduced three radial microwave switches at the mixing chamber stage inside the dilution refrigerator (see \cref{fig:Diagram}). These switches allowed us to bypass the device and measure transmission through all the possible combinations of the input and output lines. Assuming that the lines were matched to $50\, \Omega$ and there were no reflections at the switches, we calibrated out the transfer functions of the cables and amplifiers up to the magnetic shield of the device \cite{Fedorov2024}. In the second step, we fitted the measured data to a model which accounts for additional losses within the magnetic shield induced predominantly by the eccosorb filters.  This process yielded a unitary matrix that represents the device response.  

The sampling time for each scattering matrix measurement was \mbox{$\tau_s=300\,\mu\mathrm{s}$}.  As in our previous work \cite{navarathnaPhysRevLett.130.037001, Fedorov2024}, we observed the characteristic jumps between discrete output voltage states which are classified and attributed to four different quasiparticle sectors with a $K$-means classifier.  The results of the quasiparticle analysis and classification reported here exactly replicates the process we developed and described in Ref.\ \cite{Fedorov2024}.  The typical characteristic dwell times for the four quasiparticle sectors were measured as \mbox{$\tau^{\rm(qp)}_{1,2,3,4}=\{3.48 , 3.61, 4.23, 3.14\}\, \mathrm{ms}$}, which is comparable to those in our previous device \cite{Fedorov2024}.  This classification then allowed us to  compute the  circulation fidelity in each of the four quasiparticle sectors.  We then tuned the charge and flux biases to maximise the measured fidelity for one of the quasiparticle sectors.   Fixing these biases at a working point with a high circulation fidelity, we measured the scattering matrix $S$ while scanning over the drive frequency.

\begin{figure}[t!]
    \centering
\includegraphics{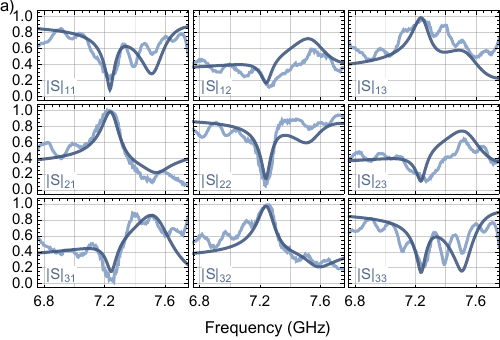}
\includegraphics{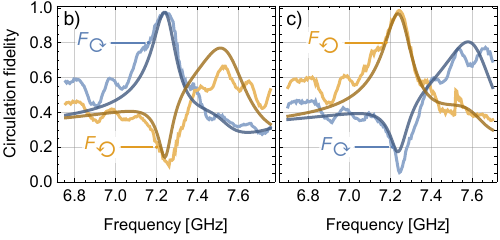}
    \caption{(a) Spectra of optimised clockwise-circulating S-matrix elements.  We note the small oscillations with period $\sim200\textrm{ MHz}$ are likely due to weak reflections in the cabling, described in \cref{append:wiggles}. (b) Clockwise and counter-clockwise fidelities for this set of S-matrix data. 
    (c) Clockwise and counter-clockwise fidelities of the S-matrix data optimally biased for counter-clockwise circulation, showing that circulation direction in the device can be changed electronically.
    }  
   \label{fig:S-matrix}
\end{figure}

The extracted scattering matrix $S$ of the device in the optimised sector is shown in \cref{fig:S-matrix}a (lighter-coloured), where we see strong nonreciprocity with
$|S_{12}|\ll|S_{21}|$, $|S_{23}|\ll|S_{32}|$, and $|S_{31}|\ll|S_{13}|$ 
around \mbox{7.25 GHz}. In addition, we observe asymmetric line shapes, which  are indicative of Fano interference, as well as noticeable background oscillations in some of the scattering matrix elements, which we attribute to weak resonances arising from back-reflection in the cabling (see \cref{append:wiggles} for more details).

We also show in  \cref{fig:S-matrix}a the results of model simulations (darker-coloured), where we used the same device parameters obtained from the spectral fit in \cref{fig:Spectrum} to compute the theoretical scattering matrix. The scattering matrix fit yields a value of the inter-waveguide capacitance $C_X=76\, \mathrm{fF}$, 
 which is reasonably close to the design value from finite-element electrostatic simulations (84 fF). The fit also returns the waveguide-loop coupling strength of $270\, \mathrm{MHz}$, which is roughly double  the value in \citet{Fedorov2024} due to a larger design ratio $C_C/C_{\Sigma}$ \cite{leOperatingPassiveOnchip2021}.

In \cref{fig:S-matrix}b, we show the clockwise and counter-clockwise circulation fidelities, ${\mathcal F}_{\,\circleright}$ and ${\mathcal F}_{\,\circleleft}$ defined in \cref{eqn:sums}, of the measured scattering spectrum in \cref{fig:S-matrix}a (lighter-coloured), along with the theoretical fidelities (darker-coloured). The peak clockwise circulation fidelity measured in \cref{fig:S-matrix}b is ${\mathcal F}_{\,\circleright}=0.97$ with a correspondingly small counter-clockwise circulation fidelity of ${\mathcal F}_{\,\circleleft}=0.12$, showing significant clockwise signal circulation. This is consistent with the theoretical simulations in \cref{fig:fideVsEJasym}, where we predict that at a Josephson-energy spread $\delta E_J = 3.2\%$ circulation fidelity ${\mathcal F}> 0.96$ is achievable for a waveguide shunt capacitance of \mbox{$C_X \sim 75\, \mathrm{fF}$}.

\begin{figure}[t!]
    \centering
    \includegraphics{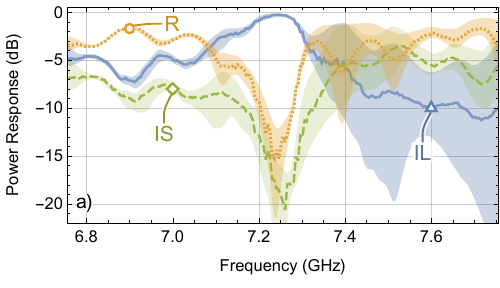}
\includegraphics{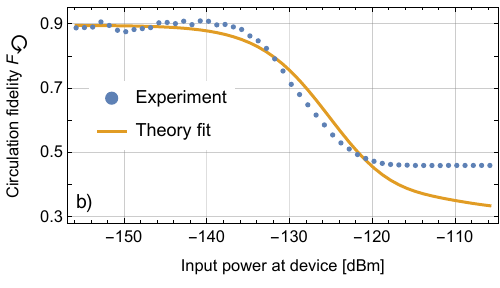}
    \caption{(a)  Clockwise circulation performance for the scattering data in \cref{fig:S-matrix}a, including the average insertion loss $\IL={\mathcal F}_{\,\circleright}^2$, the average isolation $\IS={\mathcal F}_{\,\circleleft}^2$, and the average power reflectance $\R={\mathcal R}^2$. The range of each curve indicated by a shaded region is computed from the smoothed maxima and minima over the terms in the corresponding definition in \cref{eqn:sums}. (b)  Counter-clockwise fidelity $\mathcal F_{\,\circleleft}$ at an off-resonant drive frequency $7.46\, \mathrm{GHz}$ as a function of the input power obtained from both theory and experiment, showing the saturation power of about \mbox{$P_{3{\rm dB}}\approx-126\textrm{ dBm}$}.}. 
    \label{fig:ScatteringPerformance}
\end{figure}

The direction of signal circulation in our device can be dynamically switched with the external voltage control biases. \Cref{fig:S-matrix}c shows the fidelities ${\mathcal F}_{\,\circleright}$ and ${\mathcal F}_{\,\circleleft}$ versus the drive frequency at a voltage-bias tuned for  counter-clockwise circulation; the results look very similar to those in \cref{fig:S-matrix}b, except their roles are exchanged. The peak counter-clockwise circulation fidelity  measured in \cref{fig:S-matrix}c is \mbox{${\mathcal F}_{\,\circleleft}=0.98$}, and \mbox{${\mathcal F}_{\,\circleright}=0.05$} is correspondingly small.

For completeness, we present the scattering matrices for the other quasiparticle sectors in \cref{append:quasiparticles}.  As in Ref.\ \cite{Fedorov2024}, the clockwise circulation performance in the other, unoptimised quasiparticle sectors is significantly worse than the sector reported in \cref{fig:S-matrix}a.

\subsection{Circulation performance and power dependence}

We analyse the device performance as a clockwise circulator by defining the average insertion loss $\IL$, the average isolation $\IS$, and the average power reflectance $\R$ respectively as
\begin{equation}
    \IL={\mathcal F}_{\,\circleright}^2, \hspace{0.5cm} \IS={\mathcal F}_{\,\circleleft}^2, \hspace{0.5cm} \R={\mathcal R}^2.
\end{equation}

These quantities are computed from the scattering data in \cref{fig:S-matrix}a and are shown in \cref {fig:ScatteringPerformance}a, where we find that at the resonance frequency \mbox{$7.25$ GHz}, \mbox{$\IL=0.2$ dB}, \mbox{$\IS=18$ dB}, and \mbox{$\R=-15$ dB}. In addition, \cref{fig:ScatteringPerformance}a shows  $\IL<1\, \rm dB$ over a bandwidth of $90\, \rm MHz$, while $\IS>14\, \rm dB$ over a bandwidth of $85 \, \rm MHz$. The circulator device studied here thus exhibits a ten-fold improvement in the insertion loss relative to the earlier device reported in \cite{Fedorov2024} (for which  $\IL=2\, \rm dB$, $\IS = 14\, \rm dB$, and $\R = -11\, \rm dB$ at resonance).  Based on the model predictions shown in \cref{fig:fideVsEJasym}, we attribute this to the inclusion of the shunt capacitors, which reduces the required junction fabrication precision.

Finally, to complete the characterisation of the device performance we measured the saturation power. In \cref{fig:ScatteringPerformance}b, we show the dependence of the measured counter-clockwise fidelity  on the input signal power (at an off-resonance drive frequency $7.46\, \mathrm{GHz}$), as well as including numerical simulations. The \mbox{3 dB} compression point is \mbox{$P_{3\rm dB} \approx -126 \, \rm dBm$}, which is the same as in our previous circulator device \cite{Fedorov2024}.

Microwave circulation in the device arises from   quantum interference between the ground and excited states of the junction loop.  The junction-loop energy spectrum is strongly anharmonic, and so we expect the device saturation power to correspond to the arrival of one drive photon per excited-state lifetime, \mbox{$\tau_e=(\Gamma \,|\bra{e}\hat n_a\ket{g}|^2)^{-1}$}. 
Using our fitted device parameters, we find $\Gamma \approx 270\, \mathrm{MHz}$ and $|\bra{e}\hat n_a\ket{g}|^2 \approx 0.3$, which implies a saturation power \mbox{$P_{\rm sat}\approx h f/ \tau_e = - 124\, \rm dBm $}, consistent with the experimentally measured value for \mbox{$P_{3\rm dB}$} above.

\section{Discussion and Conclusions}
In this work, we analysed an improved design for an on-chip superconducting circulator based on a three-Josephson-junction loop. The key advance over our earlier results  \cite{kochTimereversalsymmetryBreakingCircuitQEDbased2010,mullerPassiveOnChipSuperconducting2018,navarathnaPhysRevLett.130.037001,Fedorov2024} was the addition of shunt capacitors between the input-output waveguides to induce Fano interference between the scattering pathways. We showed  theoretically and experimentally that circulation in the capacitively-shunted, three-junction loop is substantially less sensitive to Josephson junction asymmetry and exhibits notable improvement in the circulation performance. 

This simple design modification increased the tolerable junction asymmetry to $\delta E_J\lesssim 3\%$, removing the need for additional fabrication post-processing to fine-tune the Josephson junctions towards higher  symmetry.  The approach may also be applicable to other  multi-component interference devices in reducing the sensitivity to typical imprecision in component fabrication.  

The device insertion loss, isolation, and return loss, which are postselected on the optimised quasiparticle sector, are comparable to the performance of commercial ferrite circulators.  The measured  bandwidth, though not as large as in commercial devices, is already sufficient for some practical applications, for example, single qubit readout \cite{Walter17RapidSingleShotDispersiveReadout}. However, the saturation power and the nonequilibrium quasiparticles, which cause random switching into and out of the high-performance sector, remain barriers to making the device practically useful. 

\begin{acknowledgments}
This work was funded through a commercial research contract with Analog Quantum Circuits (AQC) Pty.\ Ltd.  TMS and AK each declare a financial interest in AQC. 
The authors acknowledge assistance from the Centre for Microscopy and Microanalysis at the University of Queensland, and the Australian National Fabrication Facility, ANFF-Q.
\end{acknowledgments}

\appendix

\section{Scattering between capacitively coupled waveguides} \label{append:6x6WaveguideSmatDerivation}

\begin{figure}[t!]
\includegraphics[width=0.9\columnwidth]{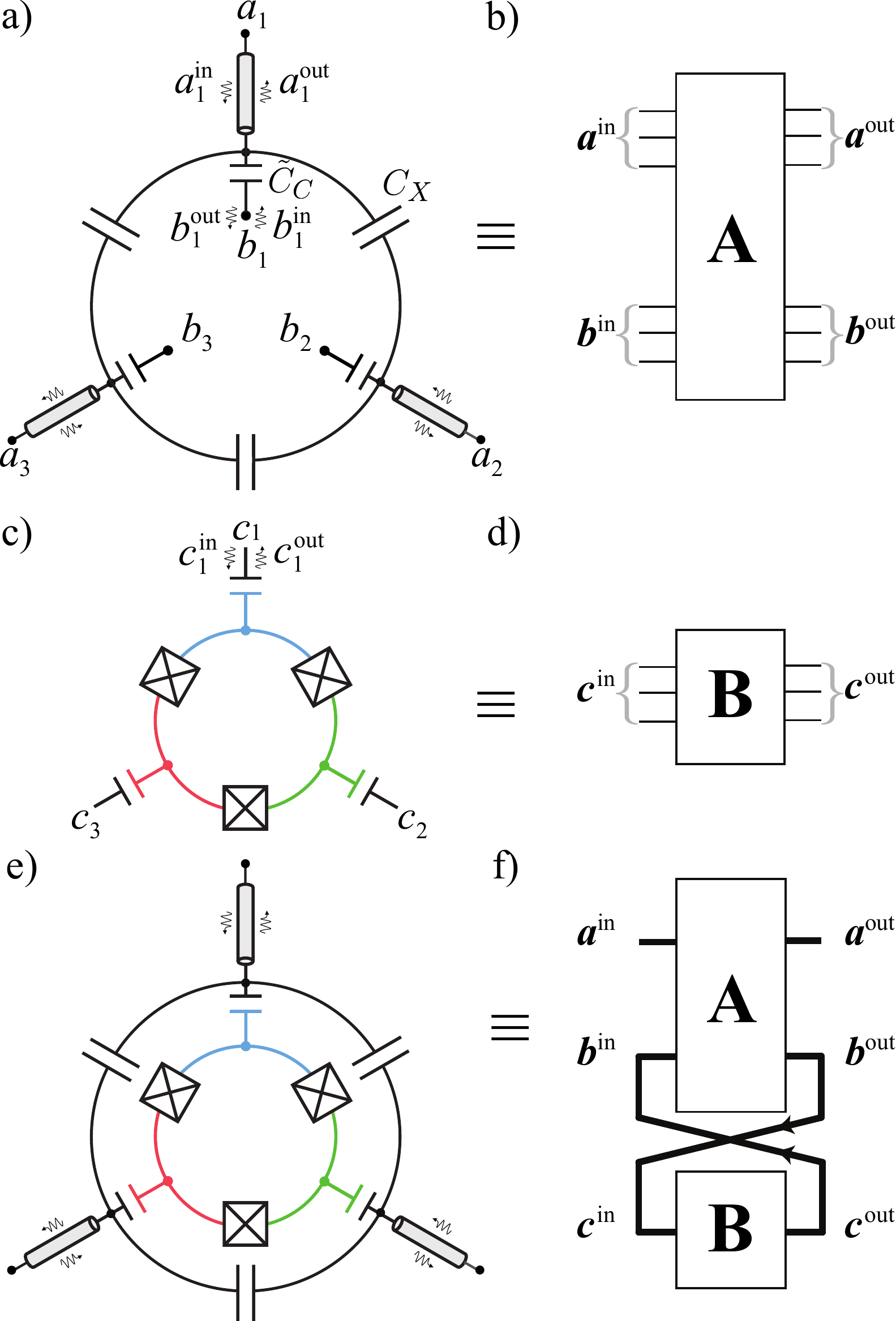}
    \caption{We describe the (a) capacitively coupled waveguides.  Formally, we take $\tilde C_C\rightarrow\infty$, to compute the transfer matrix for ${\boldsymbol{a}}^{\rm{in}}\rightarrow {\boldsymbol{b}}^{\rm{out}}$.  (b) The $6\times6$ scattering matrix $\bf{A}$ from \cref{eq:Amatrix}, relating the triplets of input and output modes of the capacitive shunts, where  \mbox{${\boldsymbol{a}}^{\rm{in / out}}=({a_1}^{\rm{in / out}},{a_2}^{\rm{in / out}},{a_3}^{\rm{in / out}})$} are the `exterior'  inputs and outputs to/from the  waveguides, and  ${\boldsymbol{b}}^{\rm{in / out}}$ are the `interior' input and output modes to/from the junction loop.   The junction loop  (c) is described by (d) a $3\times3$  junction loop  scattering matrix $\bf{B}$ relating the loop input modes, ${\boldsymbol{c}}^{\rm{in}}$, to the loop output modes, ${\boldsymbol{c}}^{\rm{out}}$, as well as the internal Hamiltonian evolution of the loop degrees of freedom. 
    The scattering matrix for the (e) waveguide-loop system from \cref{fig:circuit} is described by the SLH formalism in which $\bf{A}$ and $\bf{B}$ are coupled in a feedback loop, shown in (f), with the internal modes constrained so that ${\boldsymbol{b}}^{\rm{out}}={\boldsymbol{c}}^{\rm{in}}$ and ${\boldsymbol{b}}^{\rm{in}}={\boldsymbol{c}}^{\rm{out}}$ \cite{Josh17SLHFramework}.  The heavier lines indicate triplets of modes. Not shown here is the scattering system of the external coherent drives that couple to the exterior ports of the waveguides. }
    \label{fig:scattering}
\end{figure}

We decompose the circuit model in \cref{fig:circuit}a into a system of capacitively coupled waveguides and the three-junction loop. The waveguide system shown \cref{fig:scattering}a has three `exterior' input/output ports denoted by \mbox{$\bm{a}^{\rm in/out} = (a^{\rm in/out}_1,a^{\rm in/out}_2,a^{\rm in/out}_2)^\intercal$} that couple to external fields, and three `interior' input/output ports denoted by \mbox{$\bm{b}^{\rm in/out} = (b^{\rm in/out}_1,b^{\rm in/out}_2,b^{\rm in/out}_2)^\intercal$} that couple to the junction loop. The exterior ports couple to each other via the shunt waveguide capacitors $C_{X,j}\equiv C_X$ and couple to the interior ports via coupling capacitors $\tilde C_{C,j}\equiv \tilde C_C$. Later calculations will take the limit $\tilde C_C \to \infty$ to account for the actually galvanic connection between the exterior and interior ports. 

In the following, we derive the $6\times 6$ scattering matrix $\bf A$ for the system of waveguides in \cref{fig:scattering}a. We assume that they are capacitively connected to each other at the end points $x=0$. Following the theory of lossless semi-infinite waveguides in Ref.\ \cite{ClerkReviewNoiseMeasurementAmplification}, the voltage and current at the end point $x=0$ of the waveguide $a_j$ (or $b_j$), denoted respectively as $V_{a_j/b_j}$ and $I_{a_j/b_j}$, are given by
\begin{eqnarray}
    V_{a_j/b_j} &=& V_{a_j/b_j}^{\rm out}   + V_{a_j/b_j}^{\rm in}, \label{eq:VoltageTransLine} \\
    I_{a_j/b_j} &=& \frac{V_{a_j/b_j}^{\rm out}   - V_{a_j/b_j}^{\rm in}}{Z_{\rm wg}} , \label{eq:CurrentTransLine}
\end{eqnarray}
where $V_{a_j/b_j}^{\rm out}$ and $V_{a_j/b_j}^{\rm in}$ are the output and input voltages, 
and $Z_{\rm wg}$ is the waveguide impedance.

We apply Kirchhoff's current law at the coupling points $x=0$ of the waveguides $a_j$ and $b_j$ and find that
\begin{eqnarray}
    I_{a_j} + \frac{V_{b_j} - V_{a_j}}{Z_{\tilde{C}_j}}  + \sum_{j'\ne j} \frac{V_{a_{j'}} - V_{a_{j}}}{Z_{C_{j,j'}}}  &=& 0, \label{eq:CurrentLawAtaj} \\
    I_{b_j} + \frac{V_{a_j} - V_{b_j}}{Z_{\tilde{C}_j}}  &=& 0, \label{eq:CurrentLawAtbj} 
\end{eqnarray}
where $Z_{C_{j,j'}} = 1/i \omega_{\rm d} C_X$ and $Z_{\tilde{C}_j} = 1/i \omega_{\rm d} \tilde{C}_C$. Using Eqs.\ \eqref{eq:VoltageTransLine} and \eqref{eq:CurrentTransLine}, we rewrite Eqs.\ \eqref{eq:CurrentLawAtaj} and \eqref{eq:CurrentLawAtbj} in terms of $V^{\mathrm{out}}_{a_j/b_j}$ and $V^{\mathrm{in}}_{a_j/b_j}$
\begin{eqnarray}
    \frac{V^{\rm out}_{a_j} - V^{\rm in}_{a_j}}{Z_{\rm wg}}  + \frac{V^{\rm out}_{b_j} + V^{\rm in}_{b_j} - V^{\rm out}_{a_j} - V^{\rm in}_{a_j}}{Z_{\tilde{C}_j}} && \nonumber \\ + \sum_{j'\ne j} \frac{V^{\rm out}_{a_{j'}} + V^{\rm in}_{a_{j'}} - V^{\rm out}_{a_j} - V^{\rm in}_{a_j}}{Z_{C_{j,j'}}}  &=& 0,   \\
    \frac{V^{\rm out}_{b_j} - V^{\rm in}_{b_j}}{Z_{\rm wg}} + \frac{V^{\rm out}_{a_j} + V^{\rm in}_{a_j} - V^{\rm out}_{b_j} - V^{\rm in}_{b_j}}{Z_{\tilde{C}_j}}  &=& 0. 
\end{eqnarray}
These equations can be concisely represented  in  matrix form
\begin{equation}
   \left[  \mathds 1 + i \omega_{\rm d} Z_{\rm wg} {\bf C} \right] \left[ \begin{array}{c}
         \bm V^{\rm out}_a  \\
         \bm V^{\rm out}_b 
    \end{array} \right] =  \left[  \mathds 1 - i \omega_{\rm d} Z_{\rm wg} {\bf C} \right] \left[ \begin{array}{c}
         \bm V^{\rm in}_a  \\
         \bm V^{\rm in}_b 
    \end{array} \right], \label{eq:WaveguideInOutMatrixForm}
\end{equation}
where $\bm V_{a/b}^{\rm in/ out} = (V^{\rm in/out}_{a_1/b_1}, V^{\rm in/out}_{a_2/b_2},V^{\rm in/out}_{a_3/b_3}) $, and the full $6\times6$ capacitance matrix,
\begin{equation}
{\bf C}=    \left[ \begin{array}{cc}
       {\bm C}_X - {\bm C}_{\Sigma}  & {\bm C}_C  \\
        {\bm C}_C & - {\bm C}_C 
    \end{array} \right],
\end{equation} is defined in terms of the  $3\times3$ partial capacitance matrices ${\bm C}_\Sigma = (\tilde C_C + 2 C_X)\mathds{1}$, $    {\bm C}_C = \tilde C_C \mathds{1}$, and
\begin{equation}
    {\bm C}_X = \left[ \begin{array}{ccc}
         0 & C_X & C_X  \\
         C_X & 0 & C_X \\
         C_X & C_X & 0
    \end{array} \right].\nonumber
\end{equation}

We solve \cref{eq:WaveguideInOutMatrixForm} to obtain  the shunt capacitor scattering matrix,  
\begin{align}
    {\bf A} &= \left[  \mathds 1 + i \omega_{\rm d} Z_{\rm wg} {\bf C} \right]^{-1} . \left[  \mathds 1 - i \omega_{\rm d} Z_{\rm wg} {\bf C} \right],\label{eq:Amatrix}\\
  &\equiv \left[ \begin{array}{cc}
     \bm A_{11} & \bm A_{12}  \\
     \bm A_{21} & \bm A_{22}
\end{array} \right],\nonumber
\end{align}
which implicitly defines the $3\times3$ submatrices $\bm A_{ij}$. 
This scattering matrix is represented graphically in \cref{fig:scattering}b. 
We take the limit \mbox{$\tilde C_C \to \infty$} and define $z=\omega_{\rm d}Z_{\rm wg} C_X$, so that
\begin{eqnarray}
{\bm A}_{11}&=&{\bm A}_{22}=\frac{z}{2 i+3 z}\left[
\begin{array}{ccc}
 -2  & 1 & 1 \\
1& -2  & 1 \\
1 & 1 & -2 \\
\end{array}
\right],\nonumber\\
{\bm A}_{12}&=&{\bm A}_{21}=\frac{z}{2 i+3 z}\left[
\begin{array}{ccc}
1+\tfrac{2i}{z}  & 1 & 1 \\
1& 1+\tfrac{2i}{z}  & 1 \\
1 & 1 & 1+\tfrac{2i}{z} \\
\end{array}
\right].\nonumber
\end{eqnarray}

\begin{figure*}[]
\includegraphics[width=0.325\textwidth]{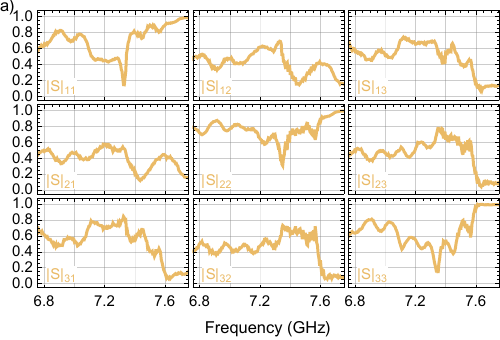}
\includegraphics[width=0.325\textwidth]{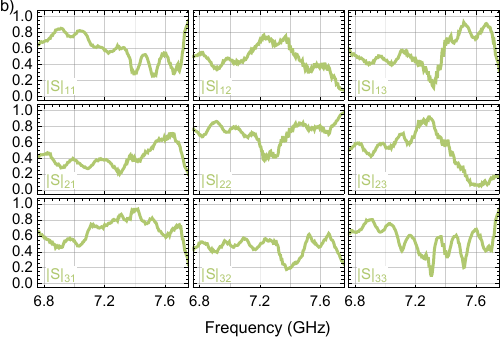}    \includegraphics[width=0.325\textwidth]{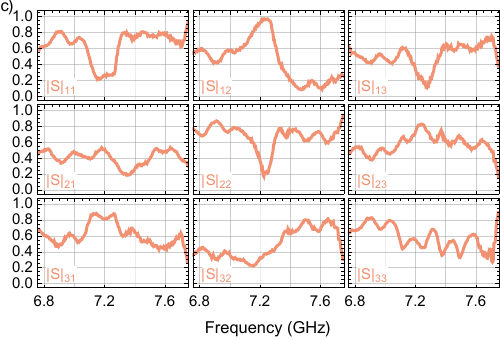}
    \caption{Scattering matrices of the other three quasiparticle sectors, obtained from the same data analysis that produced the scattering matrix in \cref{fig:S-matrix}a with high clockwise circulation.   The (clockwise) circulation performance in each of these sectors is substantially worse than in the optimised sector.
    }
   \label{fig:S-matrix_other_sectors}
\end{figure*}

\section{SLH master equation} \label{append:SLHWithFeedback}

We use the SLH formalism \cite{Josh17SLHFramework} to model the input-output network in our device. It consists of the external coherent drives, the coupled waveguides, and the junction loop. The drives are modeled as a three-port component following the source model; its SLH triple is given by
\begin{equation}
     G_{\rm d} = (\mathds 1_{3\times 3}, \hat{\bm L}_{\rm d},0),
\end{equation}
where $ \hat{\bm L}_{\rm d} = (\alpha_1 \hat{  \mathds I}, \alpha_2 \hat{\mathds I}, \alpha_3 \hat {\mathds I})^\intercal$ with $\alpha_j$ the drive amplitudes. The coupled waveguides are a six-port component without any coupling operators nor system Hamiltonian with its SLH triple of the form
\begin{equation}
     G_{\rm wg} = ({\bf A}, 0,0 ), 
\end{equation}
where $\bf A$ is given in \cref{eq:Amatrix}. 
The junction loop depicted in Figs.\ \ref{fig:scattering}c and \ref{fig:scattering}d is a three-port component represented by an SLH triple
\begin{equation}
    G_{\rm loop} = (\mathds 1_{3\times 3}, \hat{\bm{L}}_{\rm loop}, \hat H_{\rm loop}),
\end{equation}
where \mbox{$\hat {\bm{L}}_{\rm loop} = (\sqrt{\Gamma} \hat q_{1,-}, \sqrt{\Gamma} \hat q_{2,-}, \sqrt{\Gamma} \hat q_{3,-} )^\intercal$}  and $\hat H_{\rm loop}$ is given in the main text.

The total network is cascaded as
\begin{equation}
    G_{\rm tot} =  G_{\rm d} \triangleleft G_{\mathrm{w}\hookleftarrow \mathrm{l}}, \label{eq:TotalSLHTripleAp}
\end{equation}
where $G_{\mathrm{w}\hookleftarrow \mathrm{l}} = G_{\rm wg} \hookleftarrow G_{\rm loop}$ describes the feedback loop concatenation between the interior ports of $G_{\rm wg}$ and the ports of $G_{\rm loop}$, as described in Figs.\ \ref{fig:scattering}e and \ref{fig:scattering}f. We compute $G_{\mathrm{w}\hookleftarrow \mathrm{l}}$ by first cascading
$ G_{\mathrm{w} \triangleleft \mathrm{l}} = G_{\rm wg} \triangleleft (G_3 \boxplus G_{\rm loop})$, where $G_3 = (\mathds 1, 0,0)$ and  find that
\begin{equation}
  G_{\mathrm{w} \triangleleft \mathrm{l}} =  ({\bf A}, \left[ \begin{array}{c}
         0  \\
         \hat{\bm L}_{\rm loop} 
    \end{array} \right], \hat{H}_{\rm loop}).  
\end{equation}
$G_{\mathrm{w} \triangleleft \mathrm{l}}$ has six ports: its `upper' ports denoted as 1 do not involve any coupling operators and its `lower' ports denoted as 2 correspond to $\hat {\bm L}_{\rm loop}$ and are looped back to themselves. We then use the feedback rule to eliminate these internal degrees of freedom. The reduced three-port waveguide-loop SLH triple \mbox{$G_{\rm w \hookleftarrow l} = [G_{\rm w \triangleleft l}]_{2\hookleftarrow 2}$} is given by 
\begin{subequations}
\begin{eqnarray}
    S_{\rm{w \hookleftarrow l}} &=& \bm A_{11} + \bm A_{12}(\mathds 1 - \bm A_{22})^{-1} \bm A_{21}, \label{eq:WaveguideLoopSmatrix} \\
   \hat { \bm L}_{\rm w \hookleftarrow l} &=& \bm A_{12} (\mathds 1 - \bm A_{22})^{-1} \hat{\bm L_{\rm loop}}, \\
   \hat H_{\rm w \hookleftarrow l} \!&=&\! \hat H_{\rm loop} - \tfrac{i}{2} (\hat{\bm L}_{\rm loop}^\dag \bm A_{22} (\mathds 1 - \bm A_{22})^{-1} \hat{\bm L}_{\rm loop} - \rm{h.c.}). \nonumber \\
\end{eqnarray}    
\end{subequations}

Finally, the total SLH triple $G_{\rm tot}$ in \cref{eq:TotalSLHTripleAp} is
\begin{subequations}
   \begin{align}
    S_{\rm tot} &= S_{\rm w \hookleftarrow l},\label{eq:Stot} \\
    \hat{\bm L}_{\rm tot} &= \hat{\bm L}_{\rm w \hookleftarrow l} + S_{\rm w \hookleftarrow l}\hat{\bm L}_{\rm d}, \label{eq:TotalLinbladOperators} \\
    \hat H_{\rm tot} &= \hat H_{\rm w \hookleftarrow l} - \tfrac{i}{2} (\hat{\bm L}_{\rm w \hookleftarrow l}^\dag S_{\rm w \hookleftarrow l} \hat{\bm L}_{\rm d} - \rm{h.c.}). 
   \end{align}
\end{subequations}
We decompose $\hat H_{\rm tot}$ as $\hat H_{\rm tot} = \hat H_{\rm loop} + \hat H_{\rm s} + \hat H_{\rm d},$
where 
\begin{subequations}
\begin{eqnarray}
\hat H_{\rm s} &=& - \tfrac{i}{2} (\hat{\bm L}_{\rm loop}^\dag \bm A_{22} (\mathds 1 - \bm A_{22})^{-1} \hat{\bm L}_{\rm loop} - \rm{h.c.}), \quad \quad \\
\hat H_{\rm d} &=& - \tfrac{i}{2} (\hat{\bm L}_{\rm w \hookleftarrow l}^\dag S_{\rm w \hookleftarrow l} \hat{\bm L}_{\rm d} - \rm{h.c.}), 
\end{eqnarray}
\end{subequations}
which respectively describe the frequency shifts and the driving to $\hat H_{\rm loop}$. Given these, we obtain the master equation for the junction loop's density operator (in a rotating frame at the drive frequency $\omega_{\rm d}$)
\begin{equation}
    \dot \rho = - i [\hat H'_{\rm tot}, \rho] + \sum_{j=1}^3 \mathcal D[\hat L_{\mathrm{tot},j}]\rho. \label{eq:FinalSLHme}
\end{equation}
Equation \eqref{eq:TotalLinbladOperators} in fact is nothing but the input-output relation, where $\hat{\bm L}_{\rm tot} \equiv \hat{\bm a}^{\rm out}$, $\hat{\bm L}_{\rm w\hookleftarrow l}$, and $S_{\rm w\hookleftarrow l} \hat{\bm L}_{\rm d}$ represent the output fields, the system's response, and the input fields, respectively. This reproduces the master equation \cref{eq:SLHme} in the main text. We note that when $C_X =0$, one finds $\hat H_{\rm s} = 0$ and $S_{\rm w \hookleftarrow l} = \mathds 1$, which reduces \cref{eq:FinalSLHme} or \cref{eq:SLHme} to the SLH master equation used in Refs.\ \cite{mullerPassiveOnChipSuperconducting2018, leOperatingPassiveOnchip2021, navarathnaPhysRevLett.130.037001,Fedorov2024}.

\begin{figure}[t!]
    \centering
    \includegraphics[width=0.48\textwidth]{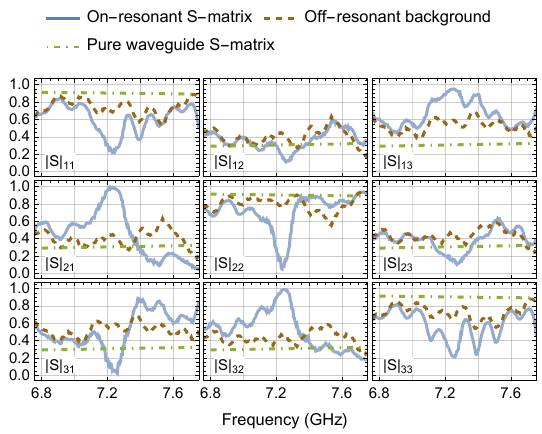}
    \caption{Comparison of the on-resonant scattering matrix data (solid blue) in \cref{fig:S-matrix}a with the off-resonant background response (dashed brown), showing very similar oscillations away from the resonant frequency $7.25\, \rm GHz$.  Also shown is the computed shunt-capacitor scattering amplitudes (dot-dashed green) from \cref{eq:Stot}, assuming \mbox{$C_X=75.7$ fF}. We attribute the period $\sim200\textrm{ MHz}$ to weak reflections in cabling, over a length of about 0.5 m.
    }
    \label{fig:background}
\end{figure}
\section{Adiabatic elimination} \label{append:AdiabaticElimination}
We consider  a semi-analytical derivation for the scattering matrix $S$ via adiabatic elimination of the SLH triple \cite{Josh17SLHFramework}, which provides us useful insights into the operation of the proposed circulator. 
In particular, we assume that the drive fields are weak so that the junction-loop system is mostly populated in its ground state \cite{mullerPassiveOnChipSuperconducting2018}. This allows us to separate the Hilbert space of the loop into a fast subspace $ \mathcal F = \{ \ket{k}, k \geq 1 \}$ that contains its excited states and a slow subspace $\mathcal S = \{ \ket{0} \}$ that contains only its ground state. We then eliminate the dynamics of the fast subspace while considering only that of the slow subspace (for more details, see Appendix B in Ref.\ \cite{leOperatingPassiveOnchip2021}). By doing so, we find that the scattering matrix $S$ within the  slow subspace is given by
\begin{equation}
     S =  ( {\mathds 1} + R_{\rm loop}).S_{\rm w \hookleftarrow l}, \label{eq:TotalScatteringMatrix}  
\end{equation}
where $R_{\rm loop}$ represents the response of the junction loop to the external coherent drives. We note that a similar expression to \cref{eq:TotalScatteringMatrix} was derived in Ref.\ \cite{Zhao19TemporalTheory} for a resonator coupled to multiple ports in the presence of a direct scattering channel between the ports.
The matrix elements of $R_{\rm loop}$ are \cite{leOperatingPassiveOnchip2021}
\begin{equation}
    (R_{\rm loop})_{ij} = - \sum_{k>0} \frac{ \langle 0 | \hat L_{\mathrm{w \hookleftarrow l},i} | k\rangle \langle k | \hat L_{\mathrm{w \hookleftarrow l},j} | 0\rangle  }{i\Delta \omega_k + \Gamma_{k}/2}, 
\end{equation}
where \mbox{$\Delta \omega_k $$= \omega_k - \omega_{\rm d}$} and 
    \mbox{$\Gamma_{k} = \langle k | \hat{\bm L}_{\rm loop}^\dag \bm{A}_{\rm s} \hat{\bm L}_{\rm loop} | k \rangle$},
with $\bm A_{\rm s} = \bm A_{22} (\mathds 1 - \bm A_{22})^{-1} $. Here $\Gamma_k$ represents the waveguide-induced decay rate as well as the frequency shift of the excited state $\ket{k}$. We numerically confirm that the S-matrices computed via the full SLH master equation \cref{eq:SLHme} and via adiabatic elimination \cref{eq:TotalScatteringMatrix} agree very well with each other.

\section{Scattering matrices of  other quasiparticle sectors}\label{append:quasiparticles}

In \cref{fig:S-matrix_other_sectors} we show the measured scattering matrices for the other quasiparticle sectors obtained from the same data analysis that yielded the scattering matrix reported in \cref{fig:S-matrix}a.  The clockwise circulation performance is evidently poor for these, compared with the best sector in \cref{fig:S-matrix}a.

\section{Background oscillations}\label{append:wiggles}

The S-matrix elements shown in \cref{fig:S-matrix}a and \cref{fig:S-matrix_other_sectors} have noticeable oscillatory responses with peak-to-trough amplitude of about 0.2, and a characteristic period of about 200 MHz.  We attribute these to weak, resonant reflections in the cabling, over a length of around 0.5 m.  

\Cref{fig:background} reproduces the resonant circulation data (solid blue) in \cref{fig:S-matrix}a, as well as similar scattering data (dashed brown) taken with the flux-bias selected to fully detune the device from the frequency band shown.  The data sets were taken several days apart.  In addition, the pure theoretical scattering matrix amplitudes for a pure shunt capacitor array, computed with \cref{eq:Stot}, are shown (dot-dashed green), which broadly match the off-resonance scattering amplitudes, apart from the small oscillations attributed to the cable back-reflections.

Both on-resonant and off-resonant data sets have similar characteristic oscillations with period $\sim200\textrm{ MHz}$, indicating that these are systematic, and reasonably stable. In some panels, far from the resonant feature around \mbox{7.25 GHz}, the oscillations match closely.  In others, the oscillations are qualitatively similar, but do not match exactly, suggesting some subtle drifts over time.

\bibliography{paper}

\end{document}